\documentclass[final,3p,times]{elsarticle}

\usepackage{amssymb}
\usepackage{graphicx}
\usepackage{amsmath}
\usepackage[english,francais]{babel}
\usepackage[T1]{fontenc}
\usepackage{array}
\usepackage{booktabs}

\journal{Comptes rendus - Physique }

\begin{document}

\begin{frontmatter}

\title{Coupled study by TEM/EELS and STM/STS of electronic properties  of C- and CN$_{x}$-nanotubes}
\author[lem,mpq]{Hong Lin}
\ead{hong.lin@univ-lyon1.fr}
\author[mpq]{J\'{e}r\^{o}me Lagoute}
\ead{jerome.lagoute@univ-paris-diderot.fr}
\author[mpq]{Vincent Repain}
\author[mpq]{Cyril Chacon}
\author[mpq]{Yann Girard}
\author[cachan]{Jean-S\'{e}bastien Lauret}
\author[lem,saragosse]{Raul~Arenal}
\author[lem]{Fran\c{c}ois Ducastelle}
\author[mpq]{Sylvie Rousset}
\author[lem]{Annick Loiseau}

\address[lem]{Laboratoire d'Etude des Microsctructures
ONERA-CNRS, BP72, 92322 Ch\^{a}tillon Cedex, France}
\address[mpq]{Laboratoire Mat\'{e}riaux et Ph\'{e}nom\`{e}nes
Quantiques, CNRS-Universit\'{e} Paris 7, \\ 10 rue Alice Domon et
L\'{e}onie Duquet, 75205 Paris Cedex 13, France}
\address[cachan]{Laboratoire de Photonique Quantique et Mol\'{e}culaire, Institut d'Alembert, Ecole Normale Sup\'{e}rieure de Cachan, 94235 Cachan Cedex, France}
\address[saragosse]{Laboratorio de microscopias avanzadas, Instituto de Nanociencia de Aragon, U. Zaragoza, c/Mariano Esquillor, 50018 Zaragoza, Spain}

\begin{abstract}
Carbon nanotubes are the focus of considerable research efforts due
to their fascinating physical properties. They provide an excellent
model system for the study of one dimensional materials and
molecular electronics. The chirality of nanotubes can lead to very
different electronic behaviours, either metallic or semiconducting.
Their electronic spectrum consists of a series of Van Hove
singularities defining a bandgap for semiconducting tubes and
molecular orbitals at the corresponding energies. A promising way to
tune the nanotubes electronic properties for future applications is
to use doping by heteroatoms. Here we report on experimental
investigation of the role of many-body interactions in nanotube
bandgaps, the visualization in direct space of the molecular
orbitals of nanotubes and the properties of nitrogen doped nanotubes
using scanning tunneling microscopy and transmission electron
microscopy as well as electron energy loss spectroscopy.

\vskip 0.5\baselineskip

\selectlanguage{francais}
\noindent{\bf R\'esum\'e} \vskip
0.5\baselineskip \noindent {\bf Etude coupl\'{e}e par TEM/EELS et STM/STS des propri\'{e}t\'{e}s structurales et \'{e}lectroniques des nanotubes C et CN$_{x}$ }

\noindent
Les nanotubes de carbone sont l'objet d'importants efforts de recherche en raison de leurs fascinantes propri\'et\'es physiques. Ils constituent un syst\`eme mod\`ele particuli\`erement int\'eressant pour l'\'etude fondamentale de mat\'eriaux \`a une dimension et pour l'\'electronique mol\'eculaire. En fonction de leur chiralit\'e, les nanotubes peuvent adopter un comportement \'electronique soit semiconducteur soit m\'etallique. Leur spectre \'electronique est domin\'e par une s\'erie de singularit\'es de Van Hove qui d\'efinit la bande interdite des tubes semiconducteurs et les orbitales mol\'eculaires situ\'ees  \`a ces \'energies. Pour contr\^{o}ler et moduler les propri\'et\'es \'electroniques des nanotubes, une voie prometteuse est d'utiliser le dopage par des h\'et\'eroatomes. Les travaux pr\'esent\'es ici portent sur l'\'etude exp\'erimentale de l'influence des interactions \`a N corps sur la valeur de la bande interdite des tubes semiconducteurs, la visualisation dans l'espace direct des orbitales mol\'eculaires des nanotubes et les propri\'et\'es des nanotubes dop\'es par l'azote en utilisant des mesures de microscopie tunnel, microscopie \'electronique \`a balayage et spectroscopie de perte d'\'energie des \'electrons.
\end{abstract}


\end{frontmatter}

\selectlanguage{english}

\section{Introduction}
Carbon nanotubes (CNTs) have attracted considerable interest for
their unique electronic properties due to their one dimensional
character \cite{NT1,Loiseau2006}. A nanotube consists in a single
graphene layer rolled up into a hollow cylinder. The many
possibilities to roll up this sheet lead to different chiralities
for the CNTs. Depending on this chirality, the resulting CNTs can be
either metallic or semiconducting. This special link between atomic
structure and electronic properties  makes the CNTs fascinating
candidates for molecular electronics. A promising way to further
tune and control the electronic structure of CNTs is to  dope them
with foreign atoms. Direct doping during the synthesis opens a
promising route  for potential applications, such as field emission,
gas sensor and negative differential resistance (NDR) devices
\cite{Terrones2004,Ewels2005,Ayala2010a,Ayala2010,Arenal2010}. In
this context, multiwall nanotubes containing nitrogen dopants
(CN$_{x}$-MWNTs) have been successfully produced by different groups
\cite{Terrones2002,Glerup2003a}. However, the as-grown
CN$_{x}$-MWNTs usually have a bamboo-like structure involving a high
density of defects, which complicates the investigation of nitrogen
configurations and nitrogen-induced effects. Compared to MWNTs,
single-wall nanotubes (SWNTs) are  better candidates, but only a few
studies on the synthesis of nitrogen-doped SWNTs (CN$_{x}$-SWNTs)
have been reported \cite{Glerup2004, Ayala2007, Lin2009}. In these
studies, chemical characterization, such as electron energy loss
spectroscopy (EELS) \cite{Glerup2004,Lin2009} or X-ray photoelectron
spectroscopy (XPS) \cite{Ayala2007}, has been employed to analyze
the C-N bonding environment, as done before on CN$_{x}$ materials,
films or multi-wall nanotubes \cite{Ewels2005,Ayala2010a}. These
studies on CN$_{x}$-SWNTs diverge, to some extent, in their
interpretation of C-N bonding. An efficient approach providing more
than chemical information is needed to understand nitrogen
incorporation and doping impact on the atomic and electronic
structure of nanotubes.

Scanning tunneling microscopy (STM) and scanning tunneling
spectroscopy (STS) are powerful tools to measure local electronic
properties of SWNTs and correlate them with their atomic structure
\cite{Wilder1998,Odom1998}. Theoretical calculations
\cite{Krasheninnikov2000, Meunier2000, Orlikowski2000} have shown
that a single defect (vacancy, pentagon, adatom) can manifest itself
as a perturbation in the carbon network in STM images and as a
modification of the local density of states (LDOS) in STS
measurements. Thus STM/STS measurements can reveal the atomic
structure and the DOS change of nanotube-based hetero-junctions
\cite{Ouyang2001a,Kim2003,Ishigami2004}, plasma-sputtered nanotubes
\cite{Buchs2007a} and other nanotube based systems. Czerw et al
\cite{Czerw2001} have reported STM/STS studies on CN$_{x}$-MWNTs.
However, studies concerning CN$_{x}$-SWNTs are still scarce.

In the case of pristine nanotubes, the pioneering experiments of
local tunneling spectroscopy on SWNTs have been interpreted within a
tight-binding single particle model
\cite{Hamada1992,Mintmire1992,Saito1992} neglecting many-body
effects. The same model was also able to reproduce optical
measurements. However it has been shown that excitons have a major
contribution to the optical transitions
\cite{Ando1997,Kane2003,Dresselhaus2007,Ando2009}. As a consequence,
the similar values of the gap of SWNTs measured either by STM ---
not expected to be sensitive to the excitons --- or optical
absorption became unclear and it is necessary to revisit the STS
study of SWNTs.

In this context, we studied by STS the electronic bandgaps of pure
C-SWNTs and solved this apparent controversy  using a model where
many-body effects are screened by the image charge induced in the
metallic substrate. In order to provide a complete picture of the
electronic structure of nanotubes as can be measured at the atomic
scale by STM, we also investigated the wavefunctions associated to
the Van Hove singularities (VHSs) which can be regarded as nanotube
molecular orbitals. These orbitals play a major role in controlling
the electronic bandstructure of carbon nanotubes upon doping or
functionalization \cite{Strano2003}. The  modification of the
electronic structure of CNTs  by nitrogen doping has also been
studied \cite{Lin2008} by coupling morphology analysis by TEM,
chemical characterization by EELS and STM/STS measurements.

\section{Experimental}

Different sorts of nanotubes were used in the present study. Using the home-made reactor at ONERA \cite{Cochon1999}, pure carbon SWNTs were synthesized in the arc discharge configuration whereas the nitrogen doped nanotubes were synthesized in the laser vaporization configuration described elsewhere \cite{Lin2008}. The synthesis products were ultrasonically dispersed in absolute ethanol for 15 min and deposited on grids for TEM studies (Philips CM20, 200 kV) of the morphology and of the diameter of nanotubes. Several tens of pure carbon nanotubes were statistically analyzed. Using a Gaussian fit, we obtained a statistical distribution of tube diameter centred on 1.4 nm with a full-width at half-maximum of 0.2 nm. A third type of tubes was synthesized in a vertical flow aerosol reactor \cite{Susi2009}. EELS experiments have been carried out in a dedicated scanning TEM (STEM) VG-HB501 (100 KV). We used a sligthly defocussed probe of a few nm in diameter, with a convergence angle of 15 mrad and a collection angle of 24 mrad. Particular attention has been devoted to contamination and damage during acquisition. STM measurements at room temperature were performed using an Omicron Nanotechnology \copyright{} variable temperature STM. STM/STS measurements were performed using an Omicron Nanotechnology \copyright{} low temperature (5K) STM operating under UHV conditions (less than $10^{-10}$ mbar). Arc and laser nanotubes were deposited from a dispersion in alcohol after sonication onto a single-crystalline Au (111) surface previously cleaned under UHV condition by repeated cycles of ion argon sputtering and annealing at 800 K. The samples were dried in air and introduced into the UHV system. CVD nanotubes were deposited \emph{in situ} onto commercial gold on glass surface previously flashed by butane flame in air. Before STM/STS measurements, the samples were outgassed for 2 hours to remove  residual molecules. All measurements were performed with tungsten tips.

\section{Electronic bandgap of CNTs on a Au(111) surface}

\subsection{Bandgap distribution in a bundle of nanotubes}

\begin{figure}[!ht]
    \centering
    \includegraphics[width=0.8\textwidth]{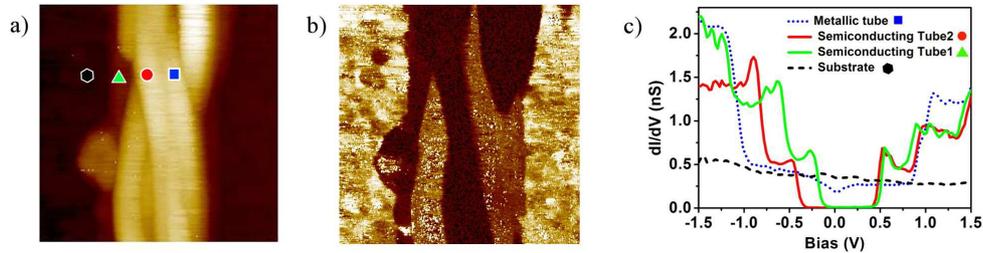}
    \caption{(Color online) STM/STS of a bundle of tubes on the substrate. (a) Topographic image. U$_{bias}$ = 1 V, I$_{setpoint}$ = 0.4 nA. (b) Differential conductance images at 20 mV. The images are 40x40 nm$^2$. The color scale corresponds to the level of DOS. At 20 mV, the semiconducting tubes have zero DOS and turn out to be darker compared to the substrate and the metallic tubes. (c) Representative dI/dV spectra taken at locations indicated by markers blue, red, green and black in image (a).}
    \label{fig:bundle}
\end{figure}

After dispersion and deposition on the Au(111) substrate, most of
the nanotubes remain arranged in bundles, allowing us to measure
bandgaps of single nanotubes, either directly in contact with the
metallic substrate, or separated from the metal by neighboring
nanotubes. In Fig.\  \ref{fig:bundle}(a), we present the topographic
image of a bundle on the Au(111) substrate. The nanotubes are
intertwined  in the bundle. The differential conductance image at 20
mV (Fig.\  \ref{fig:bundle}(b)) depicting the distribution of the
local densities of states permits us to distinguish clearly the
metallic from the semiconducting tubes in the bundle. Its color
scale corresponds to the level of the LDOS at 20 mV and reveals the
presence of semiconducting tubes in the bundle, that appear dark
whereas the metallic tubes appear with a color code closer to that
of the gold substrate.

This conductance image can be further understood by looking at the
dI/dV spectra plotted in Fig.\  \ref{fig:bundle}(c). On the gold
substrate, the differential conductance has a constant non-zero
value, as expected (black dashed curve). On the nanotubes, the
spectra are dominated by VHSs. On the metallic tube, the conductance
(blue dotted spectrum) between the two first singularities is close
to that of the substrate, except around the Fermi level where it
displays a local minimum. This pseudo-gap  \cite{Ouyang2001} can be
induced by a curvature effect \cite{Kleiner2001a} or to intertube
interactions \cite{Delaney1999}. Two spectra of semi-conducting
tubes are also shown in Fig.\ \ref{fig:bundle}(c). The spectrum of
tube 1 (green curve) exhibits a strong asymmetry with respect to the
Fermi level, which can be attributed to charge transfer from the
substrate \cite{Wilder1998,Odom1998,Venema2000} . The energy
separation between the first two singularities on each side of the
Fermi level is denoted E$^{s}_{11}$ for semiconducting tubes and
E$^{M}_{11}$ for metallic tubes, respectively. One can measure these
values at the point of maximum slope in the peak \cite{Venema2000a}
. In this way, $E^{s}_{11}$ is found to be equal to 0.70 eV (green
spectrum) for tube 1 directly contacting the substrate and to 0.91
eV for tube 2 (red spectrum) at the top of the bundle.

The renormalization of the electronic gap of a molecule due to substrate-induced charging effects in STS measurement has been already studied both theoretically and experimentally \cite{Hesper1997,Lu2004,Sau2008}. The screening effect due to the image charge in the metal was found to account for the decrease of the electronic gap of the molecules deposited on the substrate. For the nanotubes, when they are in contact with a metallic substrate, this effect should also be considered. Here the measured $E_{11}^{S}$ value of tube 1 is indeed smaller than for tube 2 and the difference  cannot be explained by the distribution in tube diameters \cite{Lin2010Nature}. This suggests that the tube-substrate distance would probably have an impact on the $E_{11}^{S}$ value measured by STS.

\subsection{Bandgap reduction due to screening by the metallic substrate}

\begin{figure}[!ht]
    \centering
\includegraphics[height=.3\textwidth]{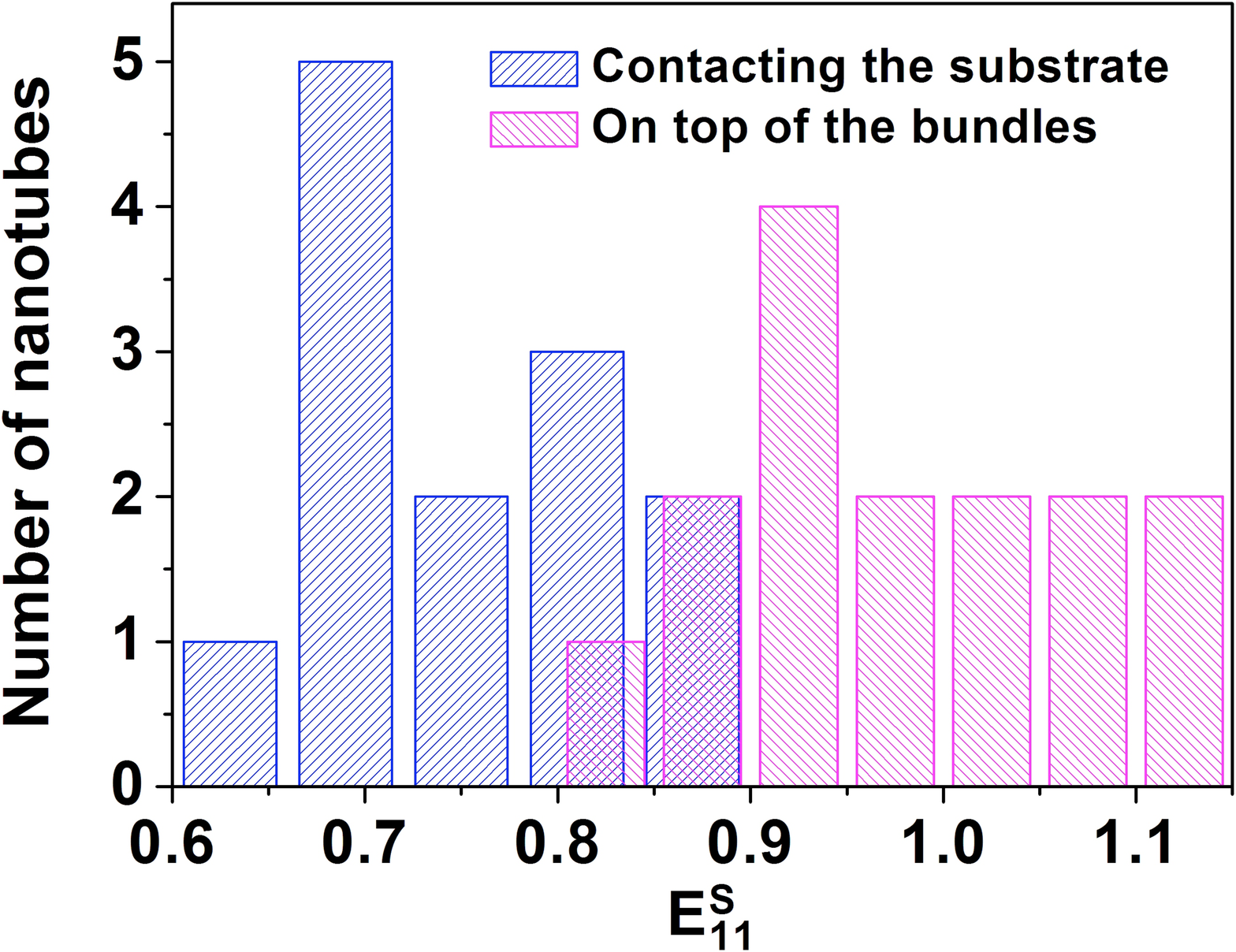}
     \caption{(Color online) Histogram of $E_{11}^{S}$ measured on the CVD sample. Blue bars: values measured on the tubes  contacting the metallic substrate, Magenta bars: values measured on top of the bundles.}
     \label{fig:SemiTKKstat}
\end{figure}

In order to verify the relation between the tube-substrate separation and the measured bandgap, we have also studied the pure carbon SWNTs synthesized by CVD. The histogram of the measured $E_{11}^{S}$ values (see Fig.\  \ref{fig:SemiTKKstat}) shows a clear right shift of $E_{11}^{S}$ measured on the top of bundles (red)
with respect to those measured on the tubes contacting the substrate (blue), in good agreement with the results obtained on the arc samples.

\begin{figure}[!ht]
    \centering
\includegraphics[height=0.3\textwidth]{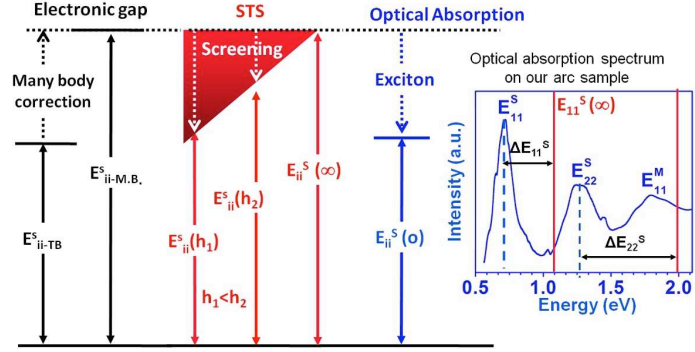}
     \caption{(Color online) Schematic illustration for the $E_{11}^{S}$transition in different situations. The
inset shows optical absorption measured on our arc nanotube sample.}
    \label{fig:explanation}
\end{figure}

The present observation can be explained as follows (see the scheme
in Fig.\ \ref{fig:explanation}). According to the literature, the
intrinsic gap of semiconducting tubes is the sum of the
single-particle gap and of the self-energy induced by the many-body
effect \cite{Wang2005,Dukovic2005}. In STS measurements, the
screening effect can be understood from an image charge model, where
the $E_{11}^{S}$ transition of a nanotube adsorbed on a metal
corresponds to the gap of the free nanotube (including
electron-electron interactions) reduced by the screening energy
resulting from the image charge in the metal $Ce^{2}/(2h_a)$, where
$h_{a}$ is the tube-substrate distance \cite{Lin2010Nature}. For the
tubes contacting the substrate ($h_a=h_1$), the many-body effect is
practically compensated by the substrate-induced screening, which
reduces significantly the gap and approaches the single-particle gap
value. As $h_{a}$ increases ($h_a=h_2$), the screening effect
decreases and leads to larger experimental STS values. The limit
$h_a\rightarrow\infty$, allows us then to estimate the unscreened
gap of nanotubes.

Experimentally, the major role played by many-body effects has been first demonstrated by optical measurements  \cite{Wang2005,Dukovic2005}. The electron-hole interaction confined in the quasi one dimension system leads to important excitonic effect in the optical absorption. At first glance, the STM experiments are not supposed to be sensitive to such an electron-hole interaction  \cite{Kane2003}. Therefore, it is instructive to compare the STS measurement with optical absorption experiments. The inset in Fig.\  \ref{fig:explanation} shows the  optical absorption spectrum measured on the same patch of arc nanotubes. The energy difference ($\Delta$E) between the optical transition and the asymptotic values $E_{ii}^{\infty}$ by STS is then expected to correspond to the exciton binding energy. In this way, we obtain $\Delta$$E_{11}^{s}\sim$ 0.4 eV, close to the value determined in previous optical measurements  \cite{Dukovic2005}.

\section{The molecular orbitals of carbon nanotubes}

In the context of controlling the electronic band structure of carbon nanotubes using doping or functionalization, the molecular orbitals of SWNTs play a major role, since they are fully involved in the transport properties and the chemical reactivity of SWNTs  \cite{Strano2003}. It has been predicted that, in defect-free semiconducting tubes, the electronic states would exhibit broken symmetry  \cite{Kane1999,Meunier1998,Meunier2000, Lambin2003a}. Local conductance measurement of SWNTs using atomically resolved STS allows us to investigate such an asymmetry at VHSs.

\subsection{Semiconducting tube: 3p+1}
\label{SC1}
\begin{figure}[!ht]
    \centering
\includegraphics[width=0.8\textwidth]{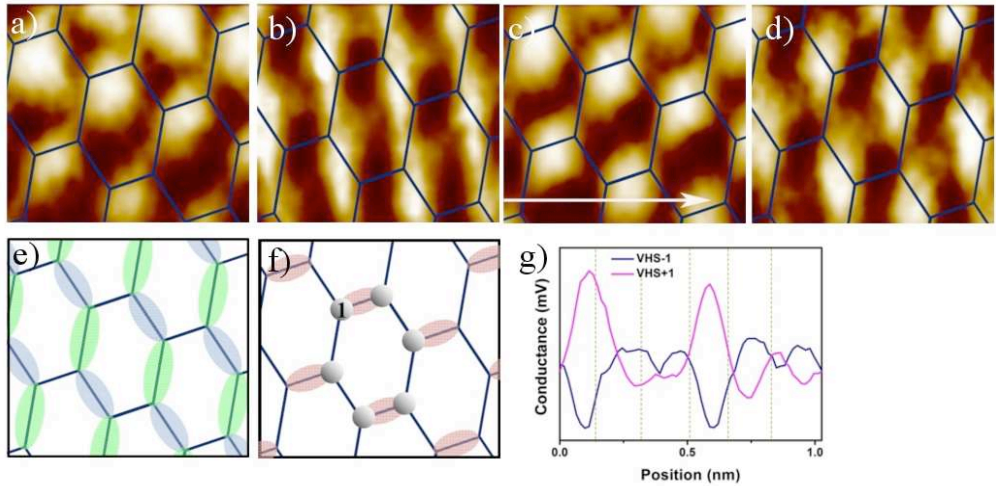}
     \caption{(Color online) Differential conductance image of the semiconducting tube at (a) VHS-2 = -0.778 V; (b) VHS-1 = -0.374V; (c) VHS+1 = 0.374 V; (d) VHS+2 = 0.778 V. The white arrow indicates the axial direction of the tube and the honeycomb lattice of the tube is drawn by blue lines. (e, f) Schematic illustration of the two wave patterns. (g) Differential conductance profiles at VHS$\pm$1 along the hexagon indicated by the blacks dots in figure (f). The starting point is the atom labeled 1 and we followed the clockwise direction. The dashed lines indicate the position of the atoms. }
 \label{fig:CITSsemi24-1}
\end{figure}

In Fig.\  \ref{fig:CITSsemi24-1}(a-d), we present the differential conductance images (dI/dV maps) in color scale of a semiconducting tube at the energies of the VHSs. The hexagonal lattice is determined on the topographic image  \cite{Lin2010} and superimposed in the images to indicate the position of carbon atoms. The significant deformation of the lattice is mostly due to the drift occurring during the long acquisition time as well as distorsions induced by the curvature of the nanotube. These conductance images show a strong anisotropy of density of states which is energy-dependent. More precisely, for the images at VHS+1 and VHS-2, one third of the C-C bonds are highlighted while the other  C-C bonds are more intense at VHS-1 and VHS+2. The four images can be summarized into two different patterns sketched in Fig.\  \ref{fig:CITSsemi24-1}(e,f). The superimposition of these two types of images, such as VHS$\pm$1, restitutes a full sixfold symmetry. In Fig.\  \ref{fig:CITSsemi24-1}(g), we plot the differential conductance profiles along an hexagon. Starting from the atom labeled 1 in Fig.\  \ref{fig:CITSsemi24-1}(f), the VHS+1 profile in the clockwise direction of the hexagon displays two maxima corresponding to higher DOS and four minima indicating lower DOS. The maxima/minima of VHS+1 are exactly at the same position as the minima/maxima of VHS-1. The conductance profile at VHS-1 has the mirror symmetry with respect to the profile at VHS+1, demonstrating clearly the complementary nature of these two patterns.

\subsection{Semiconducting tube: 3p-1}
\label{SC2} 
\begin{figure}[!ht]
    \centering
\includegraphics[width=0.8\textwidth]{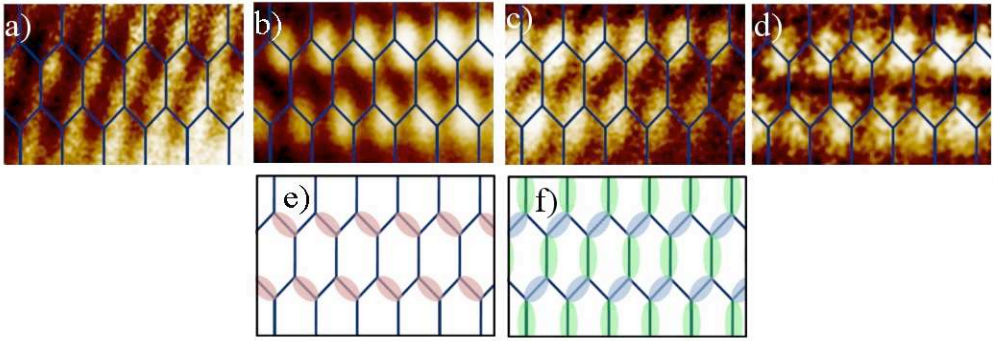}
\caption{(Color online) Differential conductance images of the
semiconducting tube 2 at (a) -0.640 V; (b) -0.302 V; (c) 0.447 V and
(d) 0.809 V. (e, f) Schematic illustration of the imaging patterns.
The tube axis is oriented in the horizontal direction.}
 \label{fig:CITSsemi74-1}
\end{figure}

Similar patterns with broken symmetry were
systematically observed on the defect-free nanotubes. Fig.\
\ref{fig:CITSsemi74-1} shows the dI/dV maps of a second type of
semiconducting tube. At VHS+1 and VHS-2, two thirds of the C-C bonds
have a maximum of LDOS. The most inclined bond with respect to the
axial orientation has maxima of electron density at VHS-1 and VHS+2,
presenting an opposite behavior  to that of the previous example. As
will be shown in the following, this behavior is the signature of
(n,m) tubes with $n-m=3p-1$ wheras the example shown in the previous
section is typical of a $n-m=3p+1$ tube.

\subsection{Tight binding analysis}
In the literature, interference patterns have been observed in STM
topographic images close to defects such as chemical impurities, cap
ends or in finite length nanotubes
\cite{Clauss1999,Furuhashi2008,Venema1999}. Differential conductance
images have also  been obtained in a finite length tube using local
differential conductance spectroscopy \cite{Lemay2001}. These
observations have been explained qualitatively by interference
effects between the Bloch waves
\cite{Kane1999,Rubio1999,Furuhashi2008}. In contrast, our results of
dI/dV maps were performed far from the extremity of the tube and
were observed systematically. Our observations are consistent with
the prediction that the electronic states would display broken
symmetry effect
\cite{Meunier1998,Clauss1999,Kane1999,Lambin2003a,Meunier2004}.

\begin{figure}[!ht]
    \centering
\includegraphics[width=0.8\textwidth]{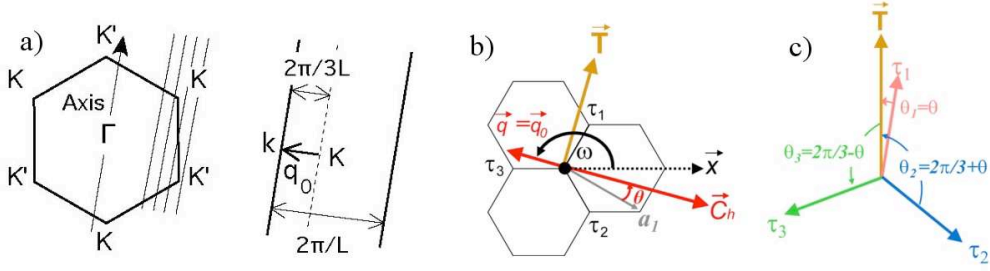}
\caption{(Color online) (a) Brillouin zone of the hexagonal graphene
lattice (left) with a zoom close to the $K$ point showing the
cutting lines parallel to the tube axis in the case of
semiconducting tube with $n-m=3p+1$ (right). (b) Relation between
$\omega$, $\vec{q}$ and chiral angle $\theta$. $\omega$ is the angle
between the $\vec{q}$ vector and the $\vec{x}$ axis and is related
to the chiral angle $\theta$. $\theta$ is defined as the smallest
angle between $\vec{a_{1}}$ vector and chiral vector $\vec{C_h}$ and
$0\leq \theta \leq \pi/6$. For the LUMO state of a $n-m=3p+1$ tube,
$\vec{q}=\vec{q_0}$, $\vec{q_0}$ being a vector of length $2\pi/3L$
pointing along the direction of the chiral vector in the negative
direction, the corresponding $\omega=\theta+\frac{5\pi}{6}$. (c)
Illustration of the angle
$\theta_\alpha=\omega-\frac{\pi}{6}-\vec{K}.\vec{\tau}{_\alpha}$ for
the LUMO state of the tubes of $n-m=3p+1$ type, $\alpha=1,2,3$.
$\theta_\alpha=\theta$, $\theta+2\pi/3$ and $\theta-2\pi/3$,
respectively. Thus, $\theta_\alpha$ represents the angle between the
vector $\vec{\tau}_{\alpha}$ and the tube axis $\vec{T}$.}
\label{fig:brillouinzone}
\end{figure}

A chemist approach in real space \cite{Lin2010} appears to be very fruitful also as we show now. We follow the classical zone-folding theory based on a $\pi$ tight binding description of the graphene band structure. The rolling up of the graphene sheet implies that the allowed $\vec{k}$ vectors belong to the so-called cutting lines parallel to the tube axis, separated by $2\pi/L$, where $L=\pi d$ is the length of the chiral vector and $d$ is the diameter of the tube. If (n,m) are the coordinates of the chiral vector, the position of the cutting lines with respect to the $K$ points depends on the value of $n-m$ modulo 3. In the case of metallic tubes, $n-m=3p$ and some cutting lines are going through the $K$ points. In the case of semiconducting tubes, $n-m=3p\pm1$, and the nearest line is at a distance $2\pi/3L$ on the left (resp. right) hand side of $K$ point when $n-m=3p+1$ (resp. $3p-1$). This depends on the specific choice of the $K$ point (see Fig.\  \ref{fig:brillouinzone}(a)), but the final result of our analysis will not. Close to the $K$ point, the linearity of the dispersion relation of graphene implies that the energy $E$ is proportional to $|\vec{k}-\vec{K}|$. For a $n-m=3p+1$ tube, as indicated in Fig.\  \ref{fig:brillouinzone}(a), the $\vec{k}$ vector of its LUMO state corresponds to $\vec{k}=\vec{K}+\vec{q}$, where $\vec{q}=\vec{q_0}$, $\vec{q_0}$ is defined as a vector of length $2\pi/3L$ pointing along the direction of the chiral vector in the negative direction. On the other side of the $K$ point, the $\vec{k}=\vec{K}+\vec{q}=\vec{K}-2\vec{q_0}$ vector corresponds to the second singularity. The Bloch eigenstates of a singularity have only a double trivial degeneracy $\pm\vec{k}$, $-\vec{k}$ being close to a $K'$ point of the Brillouin zone.

Since there are two atoms per unit cell in the graphene structure, the eigenstates $\mid\psi_{\vec{k}}\rangle$ are built from two independent Bloch states $\mid\vec{k}_{1}\rangle$ and $\mid\vec{k}_{2}\rangle$:
\begin{equation}
\label{eigenstates}
\mid \psi_{\vec{k}}\rangle=C_1\mid \vec{k_1}\rangle+C_2\mid \vec{k_2}\rangle .
\end{equation}
These two Bloch states $\mid\vec{k}_{1(2)}\rangle$ are linear combinations of atomic state $\mid\vec{n}\,\rangle$ at site $\vec{n}$ in the two sublattices.
\begin{equation}
\label{leanearcombination1}
\mid \vec{k}_{1(2)} \rangle = \sum_{\vec{n}\,\in 1(2)} e^{i\vec{k}.\vec{n}}\mid \vec{n}\, \rangle.
\end{equation}
Dropping for simplicity the $\vec{k}$ indices, the wavefunctions in real space $\psi(\vec{r})$ can be written
as:
\begin{equation}
\label{leanearcombination2}
\psi(\vec{r})=\langle\vec{r}\mid\psi\rangle =C_1\sum_{\vec{n}\in 1}e^{i
\vec{k}.\vec{n}} \phi(\vec{r}-\vec{n})+ C_2\sum_{\vec{n}\in 2}e^{i
\vec{k}.\vec{n}} \phi(\vec{r}-\vec{n})=C_1\psi_1+C_2\psi_2 \; ,
\end{equation}
where $\phi(\vec{r}-\vec{n})$ is the $\pi$ atomic orbital centered on site $\vec{n}$.
\footnote{Here we consider only the $\pi$ states.}
We can now calculate the electronic density $\rho(\vec{r})= |\psi|^2$,
\begin{equation}
\label{density4}
\rho(\vec{r})=
|C_1|^2|\psi_1|^2+|C_2|^2|\psi_2|^2+C_1C_2^\ast\psi_1\psi_2^\ast+C_1^\ast
C_2\psi_1^\ast\psi_2 .
\end{equation}
This development involves an expansion in products of the atomic wavefunctions at different sites. For $|\psi_{1(2)}|^2$:
\begin{equation}
\label{density1}
|\psi_{1(2)}|^2=\sum_{\vec{n},\vec{m}\in 1(2)}e^{i
\vec{k}.(\vec{n}-\vec{m})}
\phi(\vec{r}-\vec{n})\,\phi(\vec{r}-\vec{m}).
\end{equation}

In the first neighbor approximation, the overlap function $\phi(\vec{r}-\vec{n})\,\phi(\vec{r}-\vec{m})$ can be neglected when $n\neq m$, because first neighbour sites belong to different sublattices and the above terms reduce to the superposition of atomic densities:
\begin{equation}
\label{density2}
|\psi_{1(2)}|^2=\sum_{\vec{n}\in {1(2)}}\phi^2(\vec{r}-\vec{n}).
\end{equation}
In the same approximation, the cross term is given by:
\begin{equation}
\label{density3}
\psi_1\psi_2^\ast=\sum_{\vec{n}\in 1, \vec{m}\in 2}e^{i
\vec{k}.(\vec{n}-\vec{m})}\phi(\vec{r}-\vec{n})\,\phi(\vec{r}-\vec{m}) \; ,
\end{equation}
 where $\vec{n}$ and $\vec{m}$ are first neighbors. Close to the $K$ point, the ratio $C_1/C_2$ is given by:
\footnote{In the tight binding model, the Hamiltonian $H(\vec{k})$ has only off-diagonal non-vanishing matrix elements: $H(\vec{k})_{11}=H(\vec{k})_{22}=0$ and $H(\vec{k})_{12}=H^{\ast}(\vec{k})_{21}$. The eigenvalues of energy is given by $E=\pm\gamma_{0}|\Gamma(\vec{k})|= \pm\gamma_{0}|e^{i\vec{k}.\vec{\tau_1}}+e^{i\vec{k}.\vec{\tau_2}}+e^{i\vec{k}.\vec{\tau_3}}|$, where $\gamma_{0}$ is thet transfer integral between first neighbors. $\vec{\tau_1}$, $\vec{\tau_2}$ and $\vec{\tau_3}$ are the vectors joining one atom
to its three first neighbors, as indicated in Fig.\ \ref{fig:brillouinzone}. Close to a $K$ point, $\vec{k}=\vec{K}+\vec{q}$ and $\Gamma(\vec{K})=0$. Thus, by developing $\Gamma(\vec{k})$ ito first order in $\vec{q}$,
$\Gamma(\vec{k})=-\frac{3a_{cc}}{2}qe^{i(\omega-\frac{\pi}{6})}$, where $a_{cc}$ is the C-C nearest neighbor distance, $q$ is the module of $\vec{q}$, $\omega$ is the angle between $\vec{q}$ and the $\vec{x}$ axis, see Fig.\  \ref{fig:brillouinzone}(a). The energy $E=\pm\frac{3a_{cc}\gamma_{0}}{2}q$ depends only on the modulus of
$\vec{q}$. Since $H\psi=E\psi$, $EC_{1}=-\gamma_{0}\Gamma(\vec{k})C_{2}$, the ratio $C_1/C_2$ is then directly obtained.}
\begin{equation}
\label{sign}
 C_1/C_2=\text{sgn}(E)e^{i(\omega-\frac{\pi}{6})}.
\end{equation}
where sgn($E$)= $\pm$ 1 depends on the sign of $E$ and  $\omega$ is the angle between $\vec{q}$ and the $\vec{x}$ axis. This implies that the ratio $C_1/C_2$ just introduces a phase factor related to the direction of the $\vec{q}$ vector, \textit{i.e.\ }finally to the chiral angle $\theta$, as indicated in Fig.\  \ref{fig:brillouinzone}(b). The chiral angle $\theta$ here is precisely defined as the smallest angle between the $\vec{a_1}$ lattice vector and the chiral vector $\vec{C_h}$, so that $0\leq \theta \leq \pi/6$.
Since$ |C_1|^2=|C_2|^2$ we can normalize the density in Eq. (\ref{density4}) by fixing these quantities equal to unity. Finally, combining Equations. (\ref{density2}), (\ref{density3}) and (\ref{density4}), we  obtain the expression of the local density of states $n(\vec{r},E)$, which is the quantity measured in STS. $n(\vec{r},E)$ can be written as:
\begin{eqnarray}
\label{LDOS}
n(\vec{r},E)
&=&n(E) \times \rho(\vec{r}) \nonumber \\
&\simeq&  n(E) \times \{\sum_{\vec{n}\in
{1+2}}\phi^2(\vec{r}-\vec{n}) \nonumber\\
&+& [\text{sgn}(E)\sum_{\vec{n}\in 1, \vec{m}\in 2}e^{i[
\vec{k}.(\vec{n}-\vec{m})+\omega-\frac{\pi}{6}]}\phi(\vec{r}-\vec{n})\phi(\vec{r}-\vec{m})+
\text{complex conjugate}]\}\nonumber
\\
&=& n(E)\times\{
\rho_{\text{at}}(\vec{r})+{\rho_{\text{int}}(\vec{r})}{\}} \; ,
\end{eqnarray}
where $n(E)$ is the density of states, $\rho_{\text{at}}(\vec{r})$ is the superposition of electron atomic densities and $\rho_{\text{int}}(\vec{r})$ is the interference centered on the C-C bonds.

The breaking of the sixfold symmetry is caused by the second term $\rho_{\text{int}}(\vec{r})$. Its intensity is proportional to the nearest neighbor overlap. Then, $\vec{m}-\vec{n}$ is one of the  $\vec{\tau}{_\alpha}$ vectors between first neighbors shown in Fig.\  \ref{fig:brillouinzone}(b,c).  Keeping only the terms of lowest order, we can also replace $\vec{k}$ by $\vec{K}$, and the interference term $\rho_{\text{int}}(\vec{r})$ can be written as:
\begin{equation}
\label{interferenceterm}
\rho_{\text{int}}(\vec{r})=
2\phi(\vec{r})\phi(\vec{r}-\vec{\tau}{_\alpha} )\times \text{sgn}(E)
\cos \theta_\alpha \; ,
\end{equation}
where $\theta_\alpha=\omega-\frac{\pi}{6}-\vec{K}.\vec{\tau}{_\alpha}$.

From Eq.\  (\ref{interferenceterm}), we  see that the electronic density of the interference term is determined by the factor $\text{sgn}(E)\cos\theta_\alpha$.  To discuss the value of $\cos\theta_\alpha$, let us first consider  the LUMO state of a $n-m=3p+1$ tube. As mentioned above, the corresponding vector $\vec{q}=\vec{q_0}$, points along the negative direction of the chiral vector $\vec{C_h}$, so that $\omega=\theta+\frac{5\pi}{6}$ (see Fig.\  \ref{fig:brillouinzone}(b)). When  $\vec{\tau}{_\alpha}=\vec{\tau}_{1}$, $\vec{\tau}_{2}$, $\vec{\tau}_{3}$, $\vec{K}.\vec{\tau}{_\alpha}$ is equal to $2\pi/3$, 0 and $4\pi/3$ respectively and the corresponding values of
$\theta_\alpha=\omega-\frac{\pi}{6}-\vec{K}.\vec{\tau}{_\alpha}$ are equal to $\theta$, $\theta+2\pi/3$ and $\theta-2\pi/3$. As shown in Fig.\  \ref{fig:brillouinzone}(c), these angles are equal to the angles between the tube axis $\vec{T}$ and the C-C bonds $\vec{\tau}_{1, 2, 3}$. We have $\cos\theta_1 > 0, \: \cos\theta_{2,3}<0$. Switching to the  LUMO+1 case, the $\vec{q}$ vector should take the value $-2\vec{q}_0$. This leads to a shift of $\pi$ in $\omega$ with respect to the LUMO state and inverts the sign of $\cos\theta_\alpha$. Considering both the sign of $E$ and $\cos\theta_\alpha$, there are four possibilities,  as listed in table. \ref{tab_sign}. Depending on the sign of $\text{sgn}(E)\cos\theta_1$, basically two types of images are then expected:
\begin{table}[!ht]
\centering
\tabcolsep=6pt
\begin{tabular}{c|cccc}
   \toprule
   \textbf{} & \textbf{HOMO-1}  & \textbf{HOMO} & \textbf{LUMO} & \textbf{LUMO+1} \\
  \cmidrule(r){1-5}
  \textbf{$E$}  & - & - & + &  + \\
  \textbf{$\vec{q}$}  & $-2\vec{q}_0$ & $\vec{q}_0$ & $\vec{q}_0$ & $-2\vec{q}_0$ \\
  \textbf{$\cos\theta_1$}  & - & + & + &  - \\
  \textbf{$\cos\theta_2$} & + & - & - &  +\\
  \textbf{$\cos\theta_3$} & + & - & - &  +   \\
  \cmidrule(r){1-5}
  \textbf{$\text{sgn}(E)\cos\theta_1$}  & + & - & + &  - \\
  \textbf{$\text{sgn}(E)\cos\theta_2$}  & - & + & - &  + \\
  \textbf{$\text{sgn}(E)\cos\theta_3$}  & - & + & - &  + \\
  \cmidrule(r){1-5}
  \textbf{Image Type} & I & II & I &  II   \\
  \bottomrule
\end{tabular}
\caption{Sign of the phase factor $\text{sgn}(E)\cos\theta_\alpha$ of the interference term $\rho_{\text{int}}(\vec{r})$ for $n-m=3p+1$ tubes and the resulting interference image type.}\label{tab_sign}
\end{table}

\textbf{Image of Type I}: $\text{sgn}(E)\cos\theta_1>0$. As schematically illustrated in Fig.\  \ref{fig:CITSsemi24-1}(e), the electron density above the C-C bonds pointing close to the tube axis ($\vec{\tau}_1$) is reinforced, all the more when $\theta$ is small, \emph{i.e.} close to a zig-zag orientation ($\cos\theta\simeq1$).

\textbf{Image of Type II}: $\text{sgn}(E)\cos\theta_1<0$. In this case, when the chiral angle $\theta$ is close to zig-zag orientation $(\theta\simeq0, \: \cos\theta_{2,3}\simeq -1/2)$, the images present stripe reinforcements on the bonds perpendicular to the axis ($\vec{\tau}_{2,3}$), as indicated in Fig.\  \ref{fig:CITSsemi24-1}(f). The images of type I and type II are complementary and the superposition of both types reconstitutes the six-fold symmetry of the graphene network. As $\theta$ increases, $\omega-\frac{\pi}{6}-\vec{K}.\vec{\tau_3} \to 0 $ and the density on the corresponding bonds ($\vec{\tau}_3$ in Fig.\  \ref{fig:brillouinzone}(b,c)) gradually vanishes and the density  concentrates principally on the bonds $\vec{\tau}_2$.

To summarize, for the tubes such that $n-m=3p+1$, the LUMO and HUMO contributions are of type I and type II, respectively, presenting complementary natures, because the corresponding $\text{sign}(E)$ values are opposite. When going from the first VHS to the neighboring second VHS (same sign of $E$), the $\vec{q}$ vectors are along opposite directions, implying a shift of $\pi$ in the phase factor $\omega$, which inverts the sign of $\cos\theta_\alpha$. Therefore, the singular contribution at the second VHSs is also complementary of those of the first ones.

Considering now the other family of semiconducting tubes, where $n-m=3p-1$, the $\vec{q}$ vector of the LUMO state should be replaced by $-\vec{q}_0$, giving rise to a phase shift of $\pi$ with respect to the $n-m=3p+1$ tubes. Thus, the LUMO state presents the image type II. By analogy with the analysis in table. \ref{tab_sign}, their HOMO-1, HOMO, LUMO and LUMO+1 states correspond to the images of type II, I, II and I, respectively, and are complementary to those of the $n-m=3p+1$ tubes. These results agree perfectly with those obtained by Kane and Mele \cite{Kane1999}.

Turning back to experimental data, in Fig.\  \ref{fig:CITSsemi24-1}(a-d) (see section \ref{SC1}), we have observed exactly only two types of interference images as predicted by the tight-binding model. The axially oriented C-C bond has maximum electron density at VHS+1 and minimum density at VHS-1. This implies that the investigated tube belongs to the family $3p+1$. As for the tube shown in Fig.\  \ref{fig:CITSsemi74-1} (see section \ref{SC2}), its conductance images at VHS-1 and VHS+1 exhibit the interference pattern of Type I and Type II, respectively. According to our tight binding model, it belongs to the family $3p-1$.

\section{Nitrogen doped single-walled carbon nanotubes}

After having attained a deep understanding of the electronic properties of pure carbon nanotubes, we now study nitrogen incorporation and its doping impact on the atomic and electronic structure of nanotubes.

\subsection{Structural and chemical characterization of CN$_{x}$-SNWTs by TEM/EELS}

\begin{figure}[!ht]
    \centering
    \includegraphics[width=0.8\textwidth]{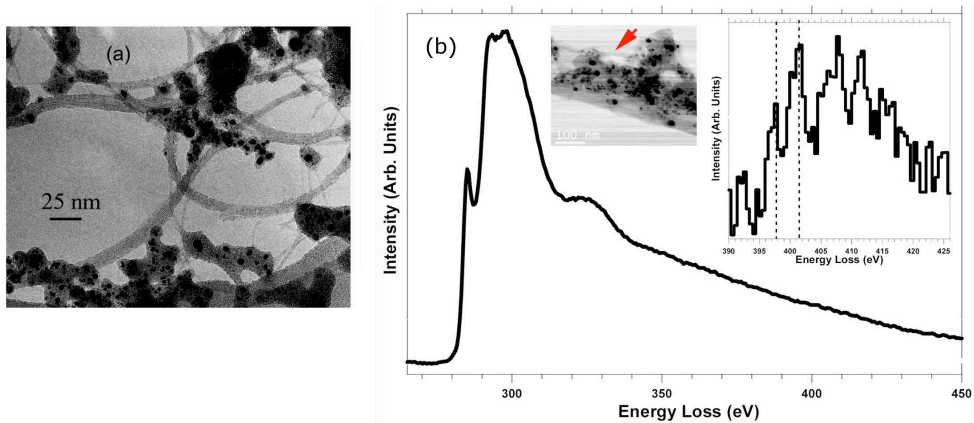}
    \caption{(Color online) (a)TEM image of dispersed CN$_{x}$-SWNTs.
    (b) Sum of 6 EEL spectra recorded on the bundle of CN$_{x}$-SWNTs displayed in
the inset (marked with an arrow). Inset: N-K edge spectrum obtained after background substraction (see the text).}
\label{TEM}
\end{figure}

Fig.\  \ref{TEM}(a) displays a TEM image of as-grown products
dispersed in solution. We observe a random network of single-walled
nanotubes, which are assembled into bundles. These bundles contain a
number of tubes varying from a few units to a few tens. The tubes
are well-crystallized, and very few defects are observed.
Local chemical characterization of as-grown nanotubes has been
previously carried out using spatially-resolved electron energy loss
spectroscopy (SREELS) of bundles of SWNTs and individual nanotubes
\cite{Lin2009}. This analysis revealed a mean nitrogen content about
1.7 atom $\%$. Fig.\  \ref{TEM}(b) shows a representative example of
the EELS spectrum recorded of those NTs. This EELS spectrum
corresponds to the sum of six EEL spectra acquired on the bundle of
NTs shown in the bright field image (acquisition conditions are
described elsewhere \cite{Lin2009}; the acquisition time was two
seconds for each  spectrum). From the shape of the C-K edge it can
be deduced that the nanotubes consist of a typical graphitic network
with sp$^{2}$-type bonding and that the tubes are very well
graphitized  \cite{Ayala2010,Lin2009}. The inset of Fig.\
\ref{TEM}(b) shows the N-K edge. Two features can be observed at
$\sim$ 398 and $\sim$ 402 eV, respectively. They could be attributed
to the $\pi$ states of the pyridine-like and graphitic-like
configuration, respectively  \cite{Ayala2010,Lin2009}. As a matter
of fact,  we observed a predominance of pyridine-like nitrogen
configuration  \cite{Lin2009}.

\subsection{Structural and electronic characterization of CN$_{x}$-SNWTs by STM/STS}

\begin{figure}
    \centering
    \includegraphics[width=90mm]{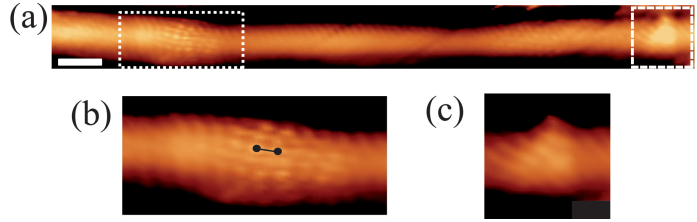}
    \caption{(Color online) (a)STM image of a SWNT containing multiple defects. V= 0.5 V, I = 200 pA, scale bar: 3 nm. The rectangle and square indicate two types of defects. (b) Magnified image of a defect with large spatial extension. The distance between black dots is equal to 0.44 nm. (c) Magnified image of a defect with small spatial extension.}
\label{Douledefects}
\end{figure}

Following this chemical analysis, STM measurements of CN$_{x}$-SWNTs have revealed the presence of defects with different characteristics. Fig.\ \ref{Douledefects}(a) displays a tube containing multiple defects. Two types of perturbation of carbon network, appearing as a local protrusion, can be readily distinguished. The rectangle indicates the first defect with large spatial extension (5$\sim$6 nm), comprising axially oriented spots with a 0.44 nm periodicity, as indicated by the black dots in the magnified image of Fig.\ \ref{Douledefects}(b). Similar defect has been previously observed in our CN$_{x}$-SWNTs and reported in  \cite{Lin2008}. We will present later on a detailed study on this kind of defect by combining STS spectroscopic data.

The second defect is indicated by the square in Fig.\
\ref{Douledefects}(a) and a magnified image in Fig.\
\ref{Douledefects}(c). It appears as a local protrusion with a
series of ``ridges'' aligned along its axial direction. Compared to
the first defect, its spatial extension is much smaller,  $\sim$1
nm. It is worth mentioning that such a signature has also been
observed several times in our CN$_{x}$-SWNTs. Even if the exact
configuration of this defect is not known, we can exclude some
cases. We first consider topological defects. The signature of a
vacancy is predicted to be a bright hillock structure
\cite{Krasheninnikov2000,Krasheninnikov2001a} and it has few in
common with the present observation. The present signature is more
similar to the presence of a pentagon in the graphitic network which
is a ring protrusion \cite{Meunier1998} or a Stone-Wales defect
exhibiting a combination of double rings protrusion with ridge-like
structure \cite{Meunier2000}. According to DFT calculations
\cite{BinZheng}, the signature of a single graphitic nitrogen is a
triangular bright spot cluster weakly sensitive to STM bias, which
seems to exclude the possibility of a single  nitrogen atom in a
substitutional site. The observation of two different defect
signatures in the same tube also indicates that these two signatures
should correspond to different atomic structures.
\begin{figure}
    \centering
    \includegraphics[width=70mm]{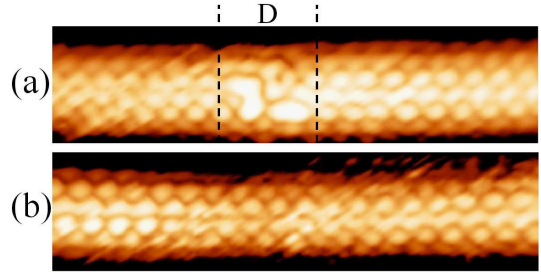}
    \caption{(Color online) STM images of a defective SWNT at 1.0 V
(a) and 1.3 V (b). I = 200 pA. A defect is localized in the D zone.
Next to the defect site, an interference pattern can be observed.}\label{tube1defaut}
\end{figure}
Fig.\  \ref{tube1defaut}(a) shows the image of a nanotube measured
at 1.0 V. The defect is localized in the middle of the image (zone
D). The difference in height between the defect and the other part
of the tube is very small, which is different from the local
protrusions presented above. It is worth noting that, next to the
defect site, the STM image exhibits an interference pattern over
long distance with a $\sqrt{3a}\times\sqrt{3a}$ periodicity ($a$ is
the graphene lattice constant). This pattern can be interpreted in
terms of electronic interference originating from large momentum
scattering of the electron wave functions at the defect site between
two nonequivalent K points, which leads to a modulation in the
electron density  \cite{Kane1999,Clauss1999}. Similar patterns have
been experimentally obtained in the vicinity of tube end caps
\cite{Furuhashi2008}, of defects created by hydrogen electron
\cite{Buchs2007} and STM tip-induced defects \cite{Berthe2007}. In
an oversimplified description, such an interference pattern is a
fingerprint of the presence of a localized defect, which can act as
an electron scatterer. In our observation, since the amplitude of
the DOS oscillation decreases with the distance, the mixture of
various electron states leads to a continuous change in the imaging
pattern \cite{Furuhashi2008}. Upon increasing the bias voltage up to
1.3 V, the perturbation area (zone D) fades out, and the tube
presents a uniform interference pattern over a distance of 10 nm,
consisting of bright spots and wavy lines. Such long range
interference patterns have been frequently observed, suggesting the
existence of a several localized defects in our CN$_{x}$-SWNTs.
\begin{figure}
    \centering
    \includegraphics[width=70mm]{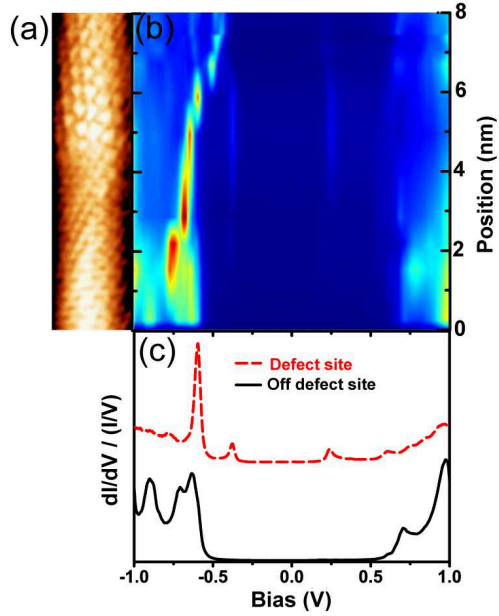}
    \caption{(Color online) The evolution of the electronic structure
    on the defective nanotube. (a) STM image of a defective tube with a local protrusion at the upper
    part of the tube. V = 1.1 V, I = 1 nA (b)
Color map of the differential conductance as a function of the bias
voltage and of the position along the defective tube in image (a).
(c) Typical STS in the defect-free zone (black curve) and on the
protrusion (red dashed curve).}
\label{STS}
\end{figure}

In order to have a deeper understanding of the defects observed in
our CN$_{x}$-SWNTs, a spectroscopic study has been undertaken at low
temperature. Fig.\  \ref{STS}(a) displays an atomically resolved STM
image of a nanotube with a protrusion with a spatial extension about
6 nm, similar to the first defect reported above. In Fig.\
\ref{STS}(b), we plot the color map of differential conductance
spectra as a function of the bias voltage and the position along the
tube, which is oriented vertically. Starting from the defect-free
zone at the bottom of the image, the tube exhibits a typical
electronic structure of a defect-free semiconducting tube. In Fig.\
\ref{STS}(c), a typical STS of this defect-free zone (black curve)
exhibits the two first VHSs, giving an electronic gap about 1.2 V.
When the tip approaches the protrusion zone, two new states appear
inside the electronic gap at around -0.37 V and 0.24 V, respectively
(dashed curve in Fig.\ \ref{STS}(c)). Referring to the literature,
paired states near the Fermi level have been attributed to the
presence of vacancy-adatom complex \cite{Kim2006} or any two close
point defects \cite{Buchs2007}. Assuming that this defect is due to
the presence of nitrogen, there are several possibilities based on
the understanding of paired states : 1) the presence of several
nitrogen atoms; 2) the combination of nitrogen with vacancy. This
observation is in line with the results presented above where the
defects observed cannot be explained by a simple substitution of a
nitrogen atom in the carbon lattice, but most likely by a complex
structure.
\subsection{Nitrogen configuration in CN$_{x}$-SWNTs}

The C-N bonding in nitrogen doped CNTs has been studied in previous works by Glerup \emph{et al.}  \cite{Glerup2004} and Ayala \emph{et al.}  \cite{Ayala2007a} who found that the nitrogen is incorporated in the walls with both graphitic and pyridine-like forms. In contrast to these analysis,
our EELS results suggest that the pyridine-like configuration is preferentially stabilized in SWNT, with respect to the graphitic one \cite{Lin2009}. Some recent theoretical studies concerning the
formation energy of pyridinic and graphitic configurations have proved the possibility of having only pyridinic configurations under some conditions: on the basis of first-principles total energy calculations, Min \emph{et al.}  \cite{Min2008} showed that the pyridine-like nitrogen configuration is indeed energetically favored as compared to graphitic configuration when two hydrogen atoms are
attached to passivate the broken carbon bond. Yang \emph{et al.}  \cite{Yang2008} showed that pyridine-like nitrogen becomes more favorable than graphite-like nitrogen as the diameter decreases and as the chiral angle increases, in agreement with  \cite{Li2008}. It is also worth noting that Ayala \emph{et al.}  \cite{Ayala2007a} have reported that pyridine-like configuration occurs at higher synthesis temperature while the graphitic substitution configuration fades away. At the highest temperature achieved in their synthesis (950$^{\circ}$), they showed a predominance of pyridine-like configuration. These results support our EELS findings.

Turning back to STM data, the protrusion with small extension
observed in our CN$_x$-SWNTs cannot be due to the presence of a
single graphitic nitrogen, whose signature is predicted to be a
triangular bright cluster. In addition, if the large protrusion in
our sample observed by STM/STS is indeed caused by nitrogen, its
electronic structure is also consistent with a non-graphitic
nitrogen configuration. It is generally believed that the graphitic
nitrogen gives rise to a donor state  \cite{Nevidomskyy2003,
Choi2000}, while we observed a double peak structure within the gap.
All these results imply that the graphitic nitrogen is not the
dominant one in our CN$_x$-SWNTs, which is in line with EELS
analysis \cite{Lin2009}. Therefore, although a chemical
characterization cannot be unambiguously performed from the STM/STS
data, we can conclude that, if nitrogen atoms are at the origin of
the defects observed, the dominant configuration is probably not a
single simple substitution.

In order to have a general idea of defect distribution and defect content in our CN$_{x}$-SWNTs, we have performed a statistical study of the atomically resolved STM images. Defects have been detected in $\sim$ 1/3 of the investigated zones (each is $\sim$ 20 nm). Assuming that all the defects observed by STM measurements are due to the presence of nitrogen, the frequency of their detection in STM images is at first sight hardly compatible with the N concentration determined from EELS, which revealed an average nitrogen concentration of $\sim$1.7 at$\%$ \cite{Lin2009}. If the distribution of nitrogen atoms on the
tube surface is homogenous, according to EELS data, we should be able to observe a defect associated with the presence of nitrogen every 4 nm (this estimation takes into account the fact that N atoms can be randomly distributed along the tube circumference and that only the upper side of tube can be detected by the STM tip). This apparent discrepancy between EELS and STM data can be explained by a heterogenous N distribution, in which case we should find some areas rich in N atoms. This leads to the question of the nature of N configurations when embedded in the graphitic network. As discussed above, nitrogen atoms are incorporated in SWNTs as complex defects, involving several nitrogen atoms. Thus, a smaller frequency of defect detection can be expected, as it is the case of our STM measurement. This also agrees with  the inhomogeneous distribution of nitrogen at the surface of tubes found in the EELS analysis. Indeed, as shown in Fig.\  \ref{Douledefects}, we observe a long range unperturbed zone ($\sim$ 20 nm), but highly perturbed
zones in some areas of the tubes.

Summarizing our results, STM data reveal the presence of different defects in our CN$_{x}$-SWNTs. These defects could be most likely made of complex structures involving one or more nitrogen and defects such as vacancies or 5/7 rings. STM and EELS results indicate that the nitrogen atoms are not regularly dispersed along the tubes. Based on this understanding, we would expect that the structure of nitrogen rich defects might vary from one defect to another, randomly dispersed along the tube.

\section{Conclusion }

In this study, we have discussed the structural and electronic properties of pristine and CN$_{x}$ single-walled nanotubes by combining TEM/EELS and STM/STS analyses. Firstly, we have evidenced the role of many-body effects played in STS and the role of the metal substrate which screens these interactions via the image charge. The comparison between STS and optical absorption measurement has allowed us to estimate the exciton binding energy of the semiconducting tubes. Secondly, we have performed a complete experimental investigation of the molecular orbitals associated with Van Hove singularities in pure semiconducting SWNTs. Using a simple tight-binding model, we have analyzed the symmetry of wavefunctions in SNWTs and demonstrated that the molecular orbitals of nanotubes at the Van Hove singularities display only  complementary behaviours. Finally, based on the analysis by TEM/EELS and STM/STS, we have tentatively proposed a model for our CN$_x$-SWNTs: the insertion of nitrogen atoms should
be associated with different structural defects, such as vacancies or pentagons and/or heptagons and the clusters of these structures disperse randomly along the tubes.

\section{Acknowledgments} This study has been supported by the European Contract STREP ``BCN'' Nanotubes 30007654-OTP25763, by a grant of CNano IdF ``SAMBA'' and by the ANR project ``CEDONA'' of the PNANO programme (ANR-07-NANO-007\_02). We gratefully acknowledge Luc
Henrard, Patrick Hermet and Bin Zheng for fruitful discussions.

%
%
%
%


%
%

\begin{thebibliography}{100}

\bibitem{NT1}
R.~Saito and G.~Dresselhaus, Physical Properties of Carbon Nanotubes, Imperial
  College Press, 1998.

\bibitem{Loiseau2006}
A.~Loiseau, P.~Launois, P.~Petit, S.~Roche, and J.~Salvetat, Understanding
  carbon nanotubes, Springer, 2006.

\bibitem{Terrones2004}
M.~Terrones, A.~Jorio, M.~Endo, A.~Rao, Y.~Kim, T.~Hayashi, H.~Terrones, J.-C.
  Charlier, G.~Dresselhaus, and M.~Dresselhaus, New direction in nanotube
  science, Materials Today \textbf{7} (2004) 30.

\bibitem{Ewels2005}
C.~P. Ewels and M.~Glerup, Nitrogen Doping in Carbon Nanotubes, Journal of
  Nanoscience and Nanotechnology \textbf{5} (2005) 1345.

\bibitem{Ayala2010a}
P.~Ayala, R.~Arenal, A.~Loiseau, A.~Rubio, and T.~Pichler, The physical and
  chemical properties of heteronanotubes, Rev. Mod. Phys. \textbf{82} (2010)
  1843.

\bibitem{Ayala2010}
P.~Ayala, R.~Arenal, M.~R\"{u}mmell, A.~Rubio, and T.~Pichler, The doping of
  carbon nanotubes with nitrogen and their potential applications, Carbon
  \textbf{48} (2010) 575.

\bibitem{Arenal2010}
R.~Arenal, X.~Blase, and A.~Loiseau, Boron-nitride and boron-carbonitride
  nanotubes: synthesis, characterization and theory, Advances in Physics
  \textbf{59} (2010) 101.

\bibitem{Terrones2002}
M.~Terrones, P.~Ajayan, F.~Banhart, X.~Blase, D.~Carroll, J.~Charlier,
  R.~Czerw, B.~Foley, N.~Grobert, R.~Kamalakaran, P.~Kohler-Redlich,
  M.~R\"{u}hle, T.~Seeger, and H.~Terrones, N-doping and coalescence of carbon
  nanotubes: synthesis and electronic properties, Applied Physics A: Materials
  Science \& Processing \textbf{74} (2002) 355.

\bibitem{Glerup2003a}
M.~Glerup, H.~Kanzow, R.~Almairac, M.~Castignolles, and P.~Bernier, Synthesis
  of multi-walled carbon nanotubes and nano-fibres using the aerosol method
  with metal-ions as the catalyst precursors, Chemical Physics Letters
  \textbf{377} (2003) 293.

\bibitem{Glerup2004}
M.~Glerup, J.~Steinmetz, D.~Samaille, O.~Stephan, S.~Enouz, A.~Loiseau,
  S.~Roth, and P.~Bernier, Synthesis of N-doped SWNT using the arc-discharge
  procedure, Chemical Physics Letters \textbf{387} (2004) 193.

\bibitem{Ayala2007}
P.~Ayala, A.~Gruneis, C.~Kramberger, M.~H. Rummeli, I.~G. Solorzano, F.~L.
  Freire, and T.~Pichler, Effects of the reaction atmosphere composition on the
  synthesis of single and multiwalled nitrogen-doped nanotubes., J. Chem. Phys.
  \textbf{127} (2007) 184709.

\bibitem{Lin2009}
H.~Lin, R.~Arenal, S.~Enouz-Vedrenne, O.~Stephan, and A.~Loiseau, Nitrogen
  Configuration in Individual CNx-SWNTs Synthesized by Laser Vaporization
  Technique, The Journal of Physical Chemistry C \textbf{113} (2009) 9509.

\bibitem{Wilder1998}
J.~W.~G. Wilder, L.~C. Venema, A.~G. Rinzler, R.~E. Smalley, and C.~Dekker,
  Electronic structure of atomically resolved carbon nanotubes, Nature
  \textbf{391} (1998) 59.

\bibitem{Odom1998}
T.~W. Odom, J.-L. Huang, P.~Kim, and C.~M. Lieber, Atomic structure and
  electronic properties of single-walled carbon nanotubes, Nature \textbf{391}
  (1998) 62.

\bibitem{Krasheninnikov2000}
A.~V. Krasheninnikov, Theoretical STM Images of Carbon Nanotubes with Atomic
  Vacancies: A Systematic Tight-Binding Study, Phys. Low-Dim. Struct
  \textbf{11/12} (2000) 1.

\bibitem{Meunier2000}
V.~Meunier and P.~Lambin, Scanning tunneling microscopy and spectroscopy of
  topological defects in carbon nanotubes, Carbon \textbf{38} (2000) 1729.

\bibitem{Orlikowski2000}
D.~Orlikowski, M.~Buongiorno~Nardelli, J.~Bernholc, and C.~Roland, Theoretical
  STM signatures and transport properties of native defects in carbon
  nanotubes, Phys. Rev. B \textbf{61} (2000) 14194.

\bibitem{Ouyang2001a}
M.~Ouyang, J.-L. Huang, C.~L. Cheung, and C.~M. Lieber, Atomically Resolved
  Single-Walled Carbon Nanotube Intramolecular Junctions, Science \textbf{291}
  (2001) 97.

\bibitem{Kim2003}
H.~Kim, J.~Lee, S.-J. Kahng, Y.-W. Son, S.~B. Lee, C.-K. Lee, J.~Ihm, and
  Y.~Kuk, Direct Observation of Localized Defect States in Semiconductor
  Nanotube Junctions, Phys. Rev. Lett. \textbf{90} (2003) 216107.

\bibitem{Ishigami2004}
M.~Ishigami, H.~J. Choi, S.~Aloni, S.~G. Louie, M.~L. Cohen, and A.~Zettl,
  Identifying Defects in Nanoscale Materials, Phys. Rev. Lett. \textbf{93}
  (2004) 196803.

\bibitem{Buchs2007a}
G.~Buchs, P.~Ruffieux, P.~Groning, and O.~Groning, Scanning tunneling
  microscopy investigations of hydrogen plasma-induced electron scattering
  centers on single-walled carbon nanotubes, Appl. Phys. Lett. \textbf{90}
  (2007) 013104.

\bibitem{Czerw2001}
R.~Czerw, M.~Terrones, J.~C. Charlier, X.~Blase, B.~Foley, R.~Kamalakaran,
  N.~Grobert, H.~Terrones, D.~Tekleab, P.~M. Ajayan, W.~Blau, M.~R\"{u}hle, and
  D.~L. Carroll, Identification of electron donor states in N-doped carbon
  nanotubes, Nano Letters \textbf{1} (2001) 457.

\bibitem{Hamada1992}
N.~Hamada, S.-i. Sawada, and A.~Oshiyama, New one-dimensional conductors:
  Graphitic microtubules, Phys. Rev. Lett. \textbf{68} (1992) 1579.

\bibitem{Mintmire1992}
J.~W. Mintmire, B.~I. Dunlap, and C.~T. White, Are fullerene tubules metallic?,
  Phys. Rev. Lett. \textbf{68} (1992) 631.

\bibitem{Saito1992}
R.~Saito, M.~Fujita, G.~Dresselhaus, and M.~S. Dresselhaus, Electronic
  structure of graphene tubules based on C60, Phys. Rev. B \textbf{46} (1992)
  1804.

\bibitem{Ando1997}
T.~Ando, Excitons in Carbon Nanotubes, Journal of the Physical Society of Japan
  \textbf{66} (1997) 1066.

\bibitem{Kane2003}
C.~L. Kane and E.~J. Mele, Ratio Problem in Single Carbon Nanotube Fluorescence
  Spectroscopy, Phys. Rev. Lett. \textbf{90} (2003) 207401.

\bibitem{Dresselhaus2007}
M.~S. Dresselhaus, G.~Dresselhaus, R.~Saito, and A.~Jorio, Exciton Photophysics
  of Carbon Nanotubes, Annual Review of Physical Chemistry \textbf{58} (2007)
  719.

\bibitem{Ando2009}
T.~Ando and U.~Seiji, Theory of Electronic States in Carbon Nanotubes, Phys.
  Stat. Sol. (c) \textbf{6} (2009) 173.

\bibitem{Strano2003}
M.~S. Strano, C.~A. Dyke, M.~L. Usrey, P.~W. Barone, M.~J. Allen, H.~Shan,
  C.~Kittrell, R.~H. Hauge, J.~M. Tour, and R.~E. Smalley, Electronic Structure
  Control of Single-Walled Carbon Nanotube Functionalization, Science
  \textbf{301} (2003) 1519.

\bibitem{Lin2008}
H.~Lin, J.~Lagoute, C.~Chacon, R.~Arenal, O.~St?han, V.~Repain, Y.~Girard,
  S.~Enouz, L.~Bresson, S.~Rousset, and A.~Loiseau, Combined STM/STS, TEM/EELS
  investigation of CNx-SWNTs, Phys. Stat. Sol. (b) \textbf{245} (2008) 1986.

\bibitem{Cochon1999}
J.-L. Cochon, J.~Gavillet, M.~L. de~la Chapelle, A.~Loiseau, M.~Ory, and
  D.~Pigache, A continuous wave CO2 laser reactor for nanotubes synthesis, in
  AIP Conf. Proc., volume 486, AIP, Kirchberg, Tirol (Austria), 1999, pp.
  237--240.

\bibitem{Susi2009}
T.~Susi, A.~G. Nasibulin, P.~Ayala, Y.~Tian, Z.~Zhu, H.~Jiang, C.~Roquelet,
  D.~Garrot, J.-S. Lauret, and E.~I. Kauppinen, High quality SWCNT synthesis in
  the presence of NH3 using a vertical flow aerosol reactor, phys. stat. sol.
  (b) \textbf{246} (2009) 2507.

\bibitem{Ouyang2001}
M.~Ouyang, J.-L. Huang, C.~L. Cheung, and C.~M. Lieber, Energy Gaps in
  ``Metallic'' Single-Walled Carbon Nanotubes, Science \textbf{292} (2001) 702.

\bibitem{Kleiner2001a}
A.~Kleiner and S.~Eggert, Curvature, hybridization, and STM images of carbon
  nanotubes, Phys. Rev. B \textbf{64} (2001) 113402.

\bibitem{Delaney1999}
P.~Delaney, H.~Joon~Choi, J.~Ihm, S.~G. Louie, and M.~L. Cohen, Broken symmetry
  and pseudogaps in ropes of carbon nanotubes, Phys. Rev. B \textbf{60} (1999)
  7899.

\bibitem{Venema2000}
L.~C. Venema, V.~Meunier, P.~Lambin, and C.~Dekker, Atomic structure of carbon
  nanotubes from scanning tunneling microscopy, Phys. Rev. B \textbf{61} (2000)
  2991.

\bibitem{Venema2000a}
L.~C. Venema, J.~W. Janssen, M.~R. Buitelaar, J.~W.~G. Wildöer, S.~G. Lemay,
  L.~P. Kouwenhoven, and C.~Dekker, Spatially resolved scanning tunneling
  spectroscopy on single-walled carbon nanotubes, Phys. Rev. B \textbf{62}
  (2000) 5238.

\bibitem{Hesper1997}
R.~Hesper, L.~H. Tjeng, and G.~A. Sawatzky, Strongly reduced band gap in a
  correlated insulator in close proximity to a metal, EPL (Europhysics Letters)
  \textbf{40} (1997) 177.

\bibitem{Lu2004}
X.~Lu, M.~Grobis, K.~H. Khoo, S.~G. Louie, and M.~F. Crommie, Charge transfer
  and screening in individual C60 molecules on metal substrates: A scanning
  tunneling spectroscopy and theoretical study, Phys. Rev. B \textbf{70} (2004)
  115418.

\bibitem{Sau2008}
J.~D. Sau, J.~B. Neaton, H.~J. Choi, S.~G. Louie, and M.~L. Cohen, Electronic
  Energy Levels of Weakly Coupled Nanostructures: C[sub 60]-Metal Interfaces,
  Phys. Rev. Lett. \textbf{101} (2008) 026804.

\bibitem{Lin2010Nature}
H.~Lin, J.~Lagoute, V.~Repain, C.~Chacon, Y.~Girard, J.-S. Lauret,
  F.~Ducastelle, A.~Loiseau, and S.~Rousset, Many-body effects in electronic
  bandgaps of carbon nanotubes measured by scanning tunnelling spectroscopy,
  Nat. Mater. \textbf{9} (2010) 235.

\bibitem{Wang2005}
F.~Wang, G.~Dukovic, L.~E. Brus, and T.~F. Heinz, The Optical Resonances in
  Carbon Nanotubes Arise from Excitons, Science \textbf{308} (2005) 838.

\bibitem{Dukovic2005}
G.~Dukovic, F.~Wang, D.~Song, M.~Y. Sfeir, T.~F. Heinz, and L.~E. Brus,
  Structural Dependence of Excitonic Optical Transitions and Band-Gap Energies
  in Carbon Nanotubes, Nano Letters \textbf{5} (2005) 2314.

\bibitem{Kane1999}
C.~L. Kane and E.~J. Mele, Broken symmetries in scanning tunneling images of
  carbon nanotubes, Phys. Rev. B \textbf{59} (1999) R12759.

\bibitem{Meunier1998}
V.~Meunier and P.~Lambin, Tight-Binding Computation of the STM Image of Carbon
  Nanotubes, Phys. Rev. Lett. \textbf{81} (1998) 5588.

\bibitem{Lambin2003a}
P.~Lambin, G.~I. Márk, V.~Meunier, and L.~P. Biró, Computation of STM images of
  carbon nanotubes, International Journal of Quantum Chemistry \textbf{95}
  (2003) 493.

\bibitem{Lin2010}
H.~Lin, J.~Lagoute, V.~Repain, C.~Chacon, Y.~Girard, F.~Ducastelle, H.~Amara,
  A.~Loiseau, P.~Hermet, L.~Henrard, and S.~Rousset, Imaging the symmetry
  breaking of molecular orbitals in single-wall carbon nanotubes, Phys. Rev. B
  \textbf{81} (2010) 235412.

\bibitem{Clauss1999}
W.~Clauss, D.~J. Bergeron, M.~Freitag, C.~L. Kane, E.~J. Mele, and A.~T.
  Johnson, Electron backscattering on single-wall carbon nanotubes observed by
  scanning tunneling microscopy, EPL (Europhysics Letters) \textbf{47} (1999)
  601.

\bibitem{Furuhashi2008}
M.~Furuhashi and T.~Komeda, Chiral Vector Determination of Carbon Nanotubes by
  Observation of Interference Patterns Near the End Cap, Phys. Rev. Lett.
  \textbf{101} (2008) 185503.

\bibitem{Venema1999}
L.~C. Venema, J.~W. Wild\&ouml;er, J.~W. Janssen, S.~J. Tans, H.~L. Tuinstra,
  L.~P. Kouwenhoven, and C.~Dekker, Imaging Electron Wave Functions of
  Quantized Energy Levels in Carbon Nanotubes, Science \textbf{283} (1999) 52.

\bibitem{Lemay2001}
S.~G. Lemay, J.~W. Janssen, M.~van~den Hout, M.~Mooij, M.~J. Bronikowski, P.~A.
  Willis, R.~E. Smalley, L.~P. Kouwenhoven, and C.~Dekker, Two-dimensional
  imaging of electronic wavefunctions in carbon nanotubes, Nature \textbf{412}
  (2001) 617.

\bibitem{Rubio1999}
A.~Rubio, D.~Sánchez-Portal, E.~Artacho, P.~Ordejón, and J.~M. Soler,
  Electronic States in a Finite Carbon Nanotube: A One-Dimensional Quantum Box,
  Phys. Rev. Lett. \textbf{82} (1999) 3520.

\bibitem{Meunier2004}
V.~Meunier and P.~Lambin, Scanning tunnelling microscopy of carbon nanotubes,
  Philosophical Transactions Of The Royal Society Of London Series
  A-Mathematical Physical And Engineering Sciences \textbf{362} (2004) 2187.

\bibitem{Krasheninnikov2001a}
A.~V. Krasheninnikov, Predicted scanning tunneling microscopy images of carbon
  nanotubes with atomic vacancies, Solid State Communications \textbf{118}
  (2001) 361.

\bibitem{BinZheng}
B.~Zheng, P.~Hermet, and L.~Henrard, personal communication, 2010.

\bibitem{Buchs2007}
G.~Buchs, A.~V. Krasheninnikov, P.~Ruffieux, P.~Groning, A.~S. Foster, R.~M.
  Nieminen, and O.~Groning, Creation of paired electron states in the gap of
  semiconducting carbon nanotubes by correlated hydrogen adsorption, New
  Journal of Physics \textbf{9} (2007) 275.

\bibitem{Berthe2007}
M.~Berthe, S.~Yoshida, Y.~Ebine, K.~Kanazawa, A.~Okada, A.~Taninaka,
  O.~Takeuchi, N.~Fukui, H.~Shinohara, S.~Suzuki, K.~Sumitomo, Y.~Kobayashi,
  B.~Grandidier, D.~Stievenard, and H.~Shigekawa, Reversible Defect Engineering
  of Single-Walled Carbon Nanotubes Using Scanning Tunneling Microscopy, Nano
  Letters \textbf{7} (2007) 3623.

\bibitem{Kim2006}
G.~Kim, B.~W. Jeong, and J.~Ihm, Deep levels in the band gap of the carbon
  nanotube with vacancy-related defects, Appl. Phys. Lett. \textbf{88} (2006)
  193107.

\bibitem{Ayala2007a}
P.~Ayala, A.~Grueneis, T.~Gemming, D.~Grimm, C.~Kramberger, M.~H. Ruemmeli,
  F.~L. Freire, H.~Kuzmany, R.~Pfeiffer, A.~Barreiro, B.~Buechner, and
  T.~Pichler, Tailoring N-doped single and double wall carbon nanotubes from a
  nondiluted carbon/nitrogen feedstock, Journal of Physical Chemistry C
  \textbf{111} (2007) 2879.

\bibitem{Min2008}
Y.-S. Min, E.~J. Bae, U.~J. Kim, E.~H. Lee, N.~Park, C.~S. Hwang, and W.~Park,
  Unusual transport characteristics of nitrogen-doped single-walled carbon
  nanotubes, Appl. Phys. Lett. \textbf{93} (2008) 043113.

\bibitem{Yang2008}
S.~H. Yang, W.~H. Shin, and J.~K. Kang, The Nature of Graphite- and
  Pyridinelike Nitrogen Configurations in Carbon Nitride Nanotubes: Dependence
  on Diameter and Helicity, Small \textbf{4} (2008) 437.

\bibitem{Li2008}
Y.~F. Li, Z.~Zhou, and L.~B. Wang, CN[sub x] nanotubes with pyridinelike
  structures: p-type semiconductors and Li storage materials, J. Chem. Phys.
  \textbf{129} (2008) 104703.

\bibitem{Nevidomskyy2003}
A.~H. Nevidomskyy, G.~Csányi, and M.~C. Payne, Chemically Active Substitutional
  Nitrogen Impurity in Carbon Nanotubes, Phys. Rev. Lett. \textbf{91} (2003)
  105502.

\bibitem{Choi2000}
H.~J. Choi, J.~Ihm, S.~G. Louie, and M.~L. Cohen, Defects, Quasibound States,
  and Quantum Conductance in Metallic Carbon Nanotubes, Phys. Rev. Lett.
  \textbf{84} (2000) 2917.


\end{thebibliography}
\end{document}